\documentclass[twocolumn]{aastex631}
\usepackage{graphicx} % Required for inserting images
\DeclareUnicodeCharacter{2212}{-}

\newcommand{\target}{KSP-OT-201803a}

\def\arcmin{\hbox{$^\prime$}}
\def\arcsec{\hbox{$^{\prime\prime}$}}

\begin{document}

\title{Photometric Study of a Distant SU UMa-type Dwarf Nova \target}
%\title{Photometric Study and Light Curves of the SU UMa-type Dwarf Nova \target}
% KSP-OT-201803a / in the field of NGC 3056
\shorttitle{S. C. Kim et al.}

\correspondingauthor{Sang Chul Kim}
\email{sckim@kasi.re.kr}

\author[0000-0001-9670-1546]{Sang Chul Kim}
\affil{Korea Astronomy and Space Science Institute, 776, Daedeok-daero, Yuseong-gu, Daejeon 34055, Republic of Korea}
\affil{Korea University of Science and Technology (UST), 217 Gajeong-ro, Yuseong-gu, Daejeon 34113, Republic of Korea}
%\email{sckim@kasi.re.kr}

\author[0000-0002-6261-1531]{Youngdae Lee}
\affil{Korea Astronomy and Space Science Institute, 776, Daedeok-daero, Yuseong-gu, Daejeon 34055, Republic of Korea}
%\email{youngdaelee@kasi.re.kr}

\author[0000-0003-4200-5064]{Dae-Sik Moon} 
\affil{David A. Dunlap Department of Astronomy and Astrophysics, University of Toronto, 50 St. George Street, Toronto, ON M5S 3H4, Canada}
%\email{daesik.moon@utoronto.ca}

\author[0000-0002-3505-3036]{Hong Soo Park} 
\affil{Korea Astronomy and Space Science Institute, 776, Daedeok-daero, Yuseong-gu, Daejeon 34055, Republic of Korea}
\affil{Korea University of Science and Technology (UST), 217 Gajeong-ro, Yuseong-gu, Daejeon 34113, Republic of Korea}
%\email{hspark@kasi.re.kr}

\author[0009-0002-8670-1096]{Dong Wook Kim}
\affil{Korea Astronomy and Space Science Institute, 776, Daedeok-daero, Yuseong-gu, Daejeon 34055, Republic of Korea}
\affil{Korea University of Science and Technology (UST), 217 Gajeong-ro, Yuseong-gu, Daejeon 34113, Republic of Korea}
%\email{dongwook@kasi.re.kr}

\shorttitle{Dwarf nova : {\target}}
\shortauthors{S. C. Kim et al.}

%==============================================================================
\begin{abstract}
We present the discovery and photometric study of a distant dwarf nova \target\ 
  using $BVI$-band observation data obtained 
  by using the KMTNet Supernova Program.
One superoutburst (SO) and three normal outbursts are identified 
  from two years of observations ($2018-2019$),
  which show that \target\ is an SU UMa-type dwarf nova (DN).
Applying the phase dispersion minimization method to the superhump data 
  during the plateau phase of the SO
  yields a superhump period of 
 $P_{\rm sh} \approx 127.4$ minutes ($=2.123$ hr $=0.08847$ days).
By stacking ten images of the quiescent phase,
  we find the quiescent phase magnitude of $V = 22.59 \pm 0.18$ mag.
The SO shows quiescent-to-peak amplitude of $4.97 \pm 0.18$ mag and duration of 16.59 days
  which are typical for the SU UMa-type DNe.
Using the superhump period-orbital period relation of SU UMa stars,
  we obtain the approximate distance of $3.7 \pm 0.3$ kpc to \target\ and
  Galactocentric distance of 9.4 kpc and height from the Galactic plane of
  $+1.2$ kpc.
Stars at these large distances may belong to the Galactic halo or thick disk, 
  which contain old, metal-poor Population II stars;
  therefore, these systems may have undergone 
  a different evolutionary process than Population I objects.
The orbital period gap of DNe spans $2.45-3.18$ hours.
Assuming that the orbital period of \target\ is a few percent shorter 
  than its superhump period, 
  \target\ may have just emerged from the period gap and recently 
  resumed mass transfer after crossing its lower boundary.
\end{abstract}

%\keywords{\uat{Dwarf Novae}{418} --- \uat{Novae}{1127} --- \uat{Time Series Analysis}{1916}}

%==============================================================================
\section{Introduction} \label{sec:intro}
Dwarf novae (DNe) are one of the cataclysmic variables (CVs) and
  close binary systems with a primary white dwarf (WD)
  and a Roche lobe-filling low mass dwarf companion 
  \citep{Warner95,Inight23}.
The companion loses and transfers mass to the primary WD
  through an accretion disk around the primary.
This causes evolution of the system toward a shorter orbital period ($P_{\rm orb}$)
  and smaller mass ratio ($q = M_2 / M_1$).
The thermal instability-induced outburst is observed as DNe
  with magnitude amplitude of $2-10$ mag and 
  the outburst durations of days to weeks \citep{Osaki96,Patterson96}.

DNe are classified into three types based on their optical light curves (LCs):
  U Geminorum (U Gem) or SS Cygni (SS Cyg)-type with 
  relatively regular outbursts and longer orbital periods,
  Z Camelopardalis (Z Cam)-type with a standstill (intermediate brightness for some time),
  and SU Ursae Majoris (SU UMa)-type with long and brighter superoutbursts (SOs)
  and short and less bright normal outbursts (NOs)
  \citep{Otula16}.
SU UMa-type DNe are further classified into ER UMa \citep{lee24}, 
  SU UMa \citep{Tampo25}, and WZ Sagittae (Sge)-types \citep{kim26}
  in the decreasing order of mass transfer rate.

As the DNe of longer orbital period ($P_{\rm orb} > 3$ hr) like
  Z Cam \citep{Simonsen14_177} and 
  U Gem \citep{Godon17} evolve,
  the donor star has a sufficiently deep convective envelope 
  and magnetic braking can efficiently remove orbital angular momentum from the binary 
  through the interaction between the donor's magnetic field and its stellar wind. 
This loss of the angular momentum causes the orbital period to decrease.
When the secondary star becomes fully convective 
  at an orbital period of $P_{\rm orb} \sim 3$ hr, magnetic braking weakens, 
  reducing the mass-transfer rate and causing the binary to become detached.
This process is thought to produce the observed period gap in DNe 
  between $\sim 2$ and $\sim 3$ hrs.

During the time the system stays in the period gap, 
  gravitational radiation \citep{Paczynski81}
  becomes the dominant mechanism driving angular-momentum loss, 
  causing the binary to shrink further.
As the Roche lobe gradually shrinks, mass transfer resumes 
  once the secondary fills its Roche lobe again
  near the lower edge of the period gap ($\sim 2$ hr).
The system continues to evolve toward the minimum orbital period
  \citep{knigge06,Schmidtobreick06,knigge11,Schreiber24},
  as the donor becomes increasingly degenerate.
After reaching the period minimum,
  the orbital period begins to increase again,
  and such a system is termed a period-bouncer.
There is, however, also an alternative idea of complexity of the surface magnetic
  field of the donor star for the explanation of the period gap
  \citep{Garraffo18}.

While the lower ($P_{\rm gap,-}$) and upper ($P_{\rm gap,+}$) edge values
  of the period gap have some ranges : $2 - 2.45$ hr and $3 - 3.18$ hr,
  respectively,
  there seems to be no consensus on the exact values,
  especially for the lower edge value.
As noted by \citet{Schreiber24}, 
  compared to the volume-limited samples,
  the SDSS samples contain fewer systems below the period gap 
  because the low mass-transfer rates in these systems make them fainter.
Consequently, low-accretion-rate CVs may be underrepresented in SDSS, 
  which identifies CVs primarily through spectroscopy. 
This observational bias may reduce the sensitivity of SDSS to 
  short-period systems with low mass-transfer rates, 
  potentially affecting the value of $P_{\rm gap,-}$.

Following the approach of the KMTNet Supernova Program (KSP) 
  \citep{lee19,lee22,lee24, kim26}, to discovering and studying DNe 
  using 24-hour photometric coverage with its 1.6 m telescopes, 
  this paper presents the discovery, photometric study, and light curve analysis of a new SU UMa-type DN, \target.
As shown in Section~\ref{sec:obs_sec}, \target\ appears to be located
  at greater distance ($\sim 3.7$ kpc, Galactocentric distance of 9.4 kpc
  and +1.2 kpc from the Galactic plane)
  than the previously known population of DNe ($\lesssim 1$ kpc, \citealt{lee19}).
At such large distances, these systems may belong to
  the Galactic halo or thick disk, which contain older, metal-poor 
  Population II stars.
Therefore, these systems may have undergone evolutionary processes 
  different from those of Population I objects.
By presenting the results for \target\ at large distance, we aim to identify and compile
  a sample of candidate Population II DNe for further investigation.

This paper is organized as follows. 
\S2 describes our observations and data reduction. 
\S3 presents the main results of our LC, 
  color-evolution, and period analyses. 
\S4 discusses the period gap, Gaia observations, and 
  the morphology of the outburst LC. 
Finally, a summary is provided in \S5.

%------------------------------------------------------------------------------
\section{Observations and Discovery} \label{sec:obs_sec}

\begin{figure}   % Fig 1 ---------------------
\epsscale{1.0}
\plotone{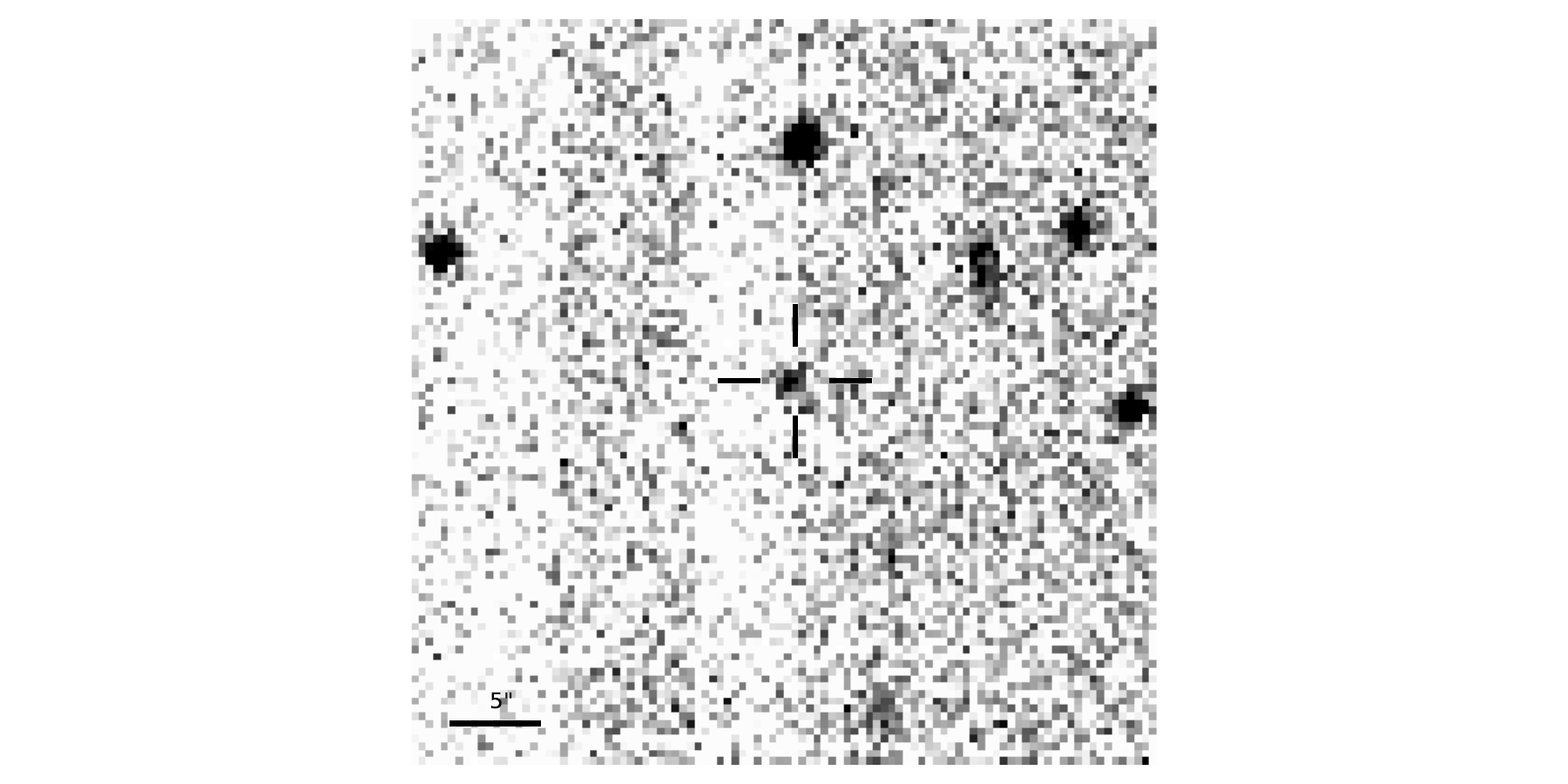}
\caption{$V$-band stacked image of ten individual exposures of 60 seconds each
  before the superoutburst during MJD $= 58170.1$ to 58174.1 days
  of \target.
North is up, and east is to the left.
Crosshair shows the position of {\target} and
  horizontal bar of 5\arcsec is shown in the bottom-left corner.
}
\label{fig:images}
\end{figure}

The coordinates of {\target} are $\alpha_{J2000}=09^{h}$ $58^{m}$ $3.25^{s}$
  ($149.513530^\circ$),
  $\delta_{J2000}= −29^\circ$ 35\arcmin~ 40.3\arcsec ($-29.594519^\circ$)
  and $(l, b)=(263.58^\circ, +19.73^\circ)$.
In this paper, for the observations, we have used the three identical 
  1.6 m telescopes of the KMTNet \citep{Kim16} installed 
  in Chile, Australia, and South Africa 
  as part of the KMTNet Supernova Program (KSP) \citep{Moon16}.
KMTNet is equipped with wide-field ($2\degr \times 2\degr$
  at a pixel scale of 0\farcs4) CCD camera and 
  an exposure time of 60 seconds are used for each of the $BVI$ filters.
{\target} was first discovered at the KMTNet in the field of 
  the lenticular galaxy NGC 3056
  by the telescope in Chile in the $B$-band
  ($B = 18.05 \pm 0.02$ mag) on 2018 Mar 29 $00^{h}$ $15^{m}$ $50.4^{s}$ 
  (UT, MJD = 58206.011), and followed by $V$ and $I$ bands :
  $V = 18.23 \pm 0.03$ mag at $00^{h}$ $17^{m}$ $16.8^{s}$ (UT) and
  $I = 17.45 \pm 0.02$ mag at $00^{h}$ $20^{m}$ $09.6^{s}$ (UT)\footnote{These 
  magnitudes are standardized values
  after applying $B$-band color correction and $I$-band magnitude shift
  which are explained in the next paragraph, but before the extinction correction.}
  on the same night.
We performed monitoring observations for \target\ 
  during MJD$=58150-58300$ and $58520-58670$ days 
  ($2018 - 2019$ years).
%------------------------------------------------------------------------------
%\subsection{Images}         \label{sec:Imag}
Figure~\ref{fig:images} shows a $V$-band image centered on \target,
  which is a stacked image produced by stacking ten individual 
  60-second exposures before the SO (MJD $= 58170.1$ to 58174.1 days).

We conducted point-spread function (PSF) photometry 
  using the DAOPHOT-II package \citep{Stetson87} and used
  the AAVSO Photometric All Sky Survey (APASS) database
  on the same target.\footnote{https://www.aavso.org/apass}
For the standardization of the magnitudes,
  we performed a $B$-band color correction of
  $\Delta B = B - B' \simeq 0.27 (B-V)$ + \textrm{offset},
  where $B$ and $B'$ are color-corrected $B$-band magnitudes 
  in the standard Johnson system 
  and uncorrected KNTNet $B$-band magnitudes, respectively \citep{Par17}.
The AAVSO photometry is provided with the $BVi$-bands,
  therefore we applied another correction of $I = i - 0.4$ mag \citep[see][]{Par17}.

We have used the Galactic dust extinction map of \citet{Green19}
  toward the position of \target.
Using equation (7) of \citet{Mennickent99} to convert 
  the superhump period $P_{\rm sh} = 0.08847$ days of \target\
  (see Section~\ref{sec:period} below) to the orbital period, 
  we obtain $P_{\rm orb} \sim 0.08444 (93)$ days\footnote{Values in parentheses 
  represent errors in the last digits.}
  ($=2.03$ hr).
The relation between the SO peak magnitude and the orbital period
  given by \citet{Patterson11} (their equation (4)) yields 
  the peak absolute magnitude of $M_{V,\rm peak} \sim 4.55 \pm 0.18$ mag
  for the SO peak of \target.
This gives an extinction-uncorrected distance modulus of 
  $(V-M_V) \sim 17.62(\pm0.02) - 4.55(\pm0.18) \approx 13.07 \pm 0.18$ mag
  (see Section~\ref{sec:LC} below for the peak magnitude of the SO)
  and an approximate distance of $\sim 4.1 \pm 0.4$ kpc.
Using this rough distance estimate, 
  the reddening is estimated to be $E(B-V) \approx 0.07$ mag and
  $E(V-I) \approx 0.09$ mag. 
Applying this reddening to the above distance estimate finally gives
  an extinction-corrected distance of  
  $d \sim 3.7 \pm 0.3$ kpc.\footnote{The corresponding Galactocentric distance
  becomes $R_{GC} = \sqrt{d^2 + R_0^2 - 2dR_0 \; cos(l) \; cos(b)}$ 
  = 9.4 $\pm$ 0.2 kpc and the height from the Galactic plane is
  $d \; sin(b)=+1.2 \pm 0.1$ kpc \citep{lee19}, assuming
  $R_0 = 8.3\pm0.2$ kpc  for the distance of the Sun from the Galactic center 
  \citep{Reid14}.}
The magnitudes given in this paper are not extinction corrected values.
Although we searched the Gaia DR3 catalog \citep{Gaia23}, 
  we could not obtain a distance estimate for \target\ 
  (see Section~\ref{sec:GAIA}).

%------------------------------------------------------------------------------
\section{Results} \label{sec:Resul} % Light Curve Analysis

\subsection{Light Curve}    \label{sec:LC}
Figure~\ref{fig:fulllc} (a) shows the light curve (LC) of \target\
  for all the observed duration of 2018 (before MJD = 58300 days) and
  2019 (after MJD = 58500 days) with a seasonal gap between them
  and zoomed-in details of the superoutburst (SO) and
  three normal outbursts (NOs: N1, N2, and N3) are shown
  in panels, (b), (c), (d), and (e), respectively.
Each of the four blue star symbols in panel (a) are the photometry results 
  from 10 stacked images (each reaching the exposure time of 600 seconds).
While we increased the tolerance range to include data
  with magnitude errors $< 0.2$ mag,
  some single-image photometry results still show values similar to
  those from the stacked-image photometry, especially at 2018 observations.
\target\ can be faintly seen in some of the single exposure images,
  although not all images show this behavior.
We have fitted the SO with order 8 polynomial function and
  shown as dashed line in panel (b).
For the three NOs,
  data for N1 and N3 are relatively sparse compared to N2.
For N2 we fitted a polynomial function with order 4 (panel (d))
  and overplotted this function in panel (c) (N1) and panel (e) (N3)
  for comparison among the NOs.
N1 appears to show a less luminous peak than N2,
  while N3 exhibits a LC shape quite similar to that of N2
  based on the small number of data points for N3.
However, these comparisons should be treated with caution
  because the numbers of data points for N1 and N3 are limited.

% SO - Physical Parameters:
From the 8th-order polynomial fit to the SO data,
  we find that the luminosity peak reached $V = 17.62 \pm 0.02$ mag at MJD $=58206.4605$ days.
After the peak, the plateau phase continued until MJD $=58218.7665$ days,
  giving a plateau duration of $\sim 12.3$ days.
We obtained quiescent phase photometry at four epochs
  and are shown in Table~\ref{tab:quiescence_table} and
  in Figure~\ref{fig:fulllc} (a) as blue star symbols.
The mean magnitudes of the quiescent phase are $B = 22.88 \pm 0.14$,
  $V = 22.59 \pm 0.18$, and $I = 22.18 \pm 0.14$ mag,
  and shown as blue horizontal dotted lines in panels (a) to (e) of 
  Figure~\ref{fig:fulllc}.
Since we have adopted magnitude errors to be $< 0.2$ mag 
  for the single image photometry, 
  the quiescent phase photometry from 10 stacked images shows
  magnitudes similar to those measured from single images  
  near MJD $\sim 58172$ and 58252 days,
  when the Moon was near its new phase and seeing conditions were good.
This is the reason why the target can be seen faintly in some of the single
  images during these time spans.
The magnitude amplitude of the SO between the peak and quiescent-phase is
  $4.97 \pm 0.18$ mag, 
  rising and decline rates are 4.54 and 1.05 mag day$^{-1}$, and
  the duration is 16.59 days (from MJD $=58204.2568$ to 58220.8514 days).

The polynomial function fitting for the N2 gives
  the peak magnitude of $V \approx 18.13$ mag at MJD $= 58600.1942$ days.
The N2 amplitude is $4.46 \pm 0.18$ mag, 
  rising and decline rates are 3.71 and 1.50 mag day$^{-1}$, and
  the duration is 4.18 days (from MJD $=58598.9930$ to 58603.1772 days).
Since we did not attempt to fit N1 and N3 due to insufficient data,
  the parameters of N2 are used to fit N1 and N3 
  solely to determine their peak MJDs, which are 58276.8942 and 
  58649.1942 days, respectively.

%------------------------------------------------------------------------------
%\subsection{SU UMa-Type Identification}  \label{sec:Type}
The SOs, NOs, and superhumps during the SO plateau indicate that \target\ is 
  an SU UMa-type DN \citep[e.g.,][]{Kato09_S395}.
\citet{Otula16} conducted a comprehensive statistical analysis of 653 DNe.
Their Figure 13 shows a strong correlation between SO and NO amplitudes
  ($A_{\rm SO}$ vs. $A_{\rm NO}$), 
  placing \target\ well within the SU UMa-type DNe. 
Their Figures 15–17 present distributions of durations, amplitudes, and 
  orbital periods for NOs and SOs.
The corresponding parameters of \target\ fall well within the ranges 
  observed in SU UMa-type DNe, regardless of the correlation strength.

%------------------------------------------------------------------------------
\subsection{Color Evolution}  \label{sec:Color}
\begin{figure*}   % Fig 2 ----------------------------------------------
\epsscale{1.0}
\plotone{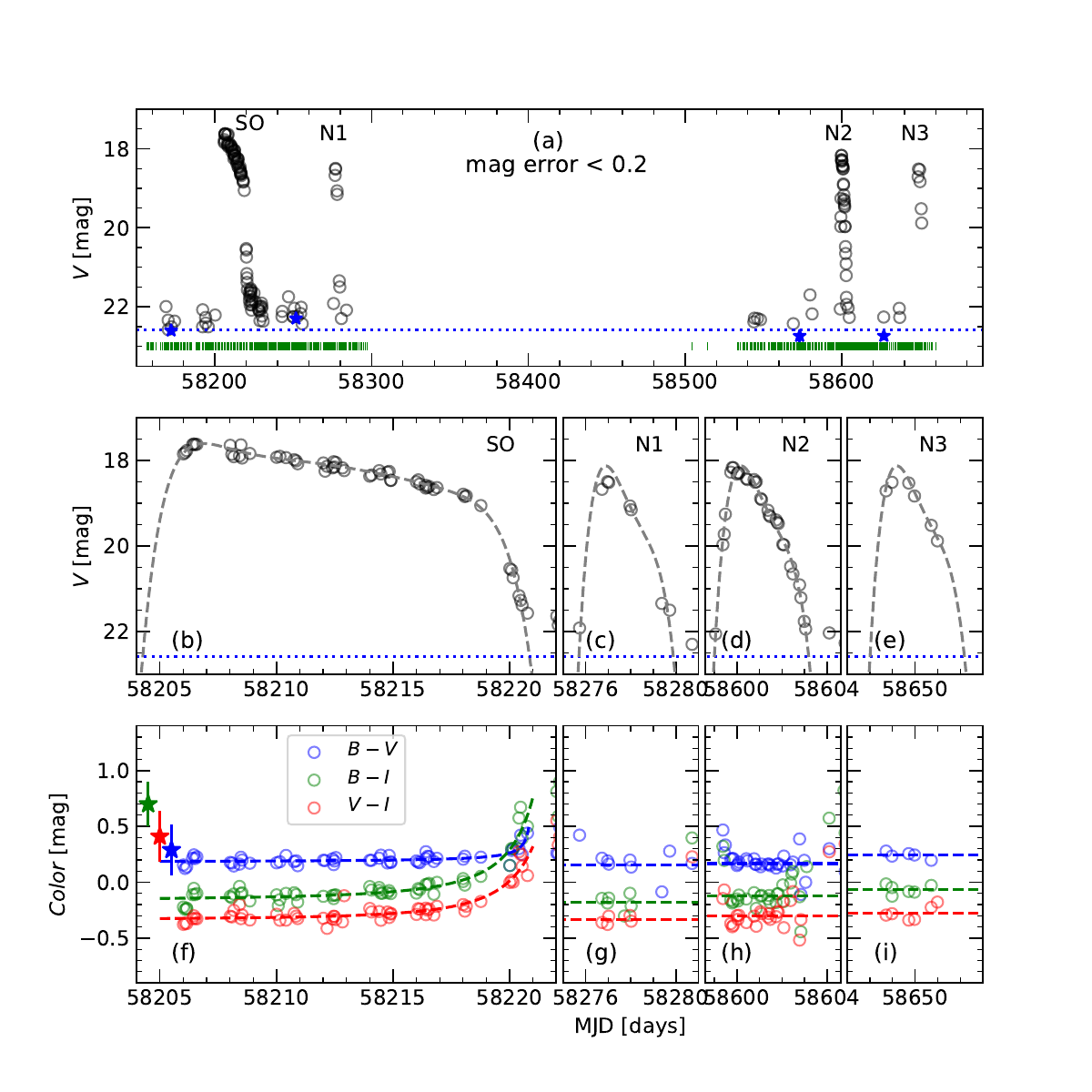}
\caption{(a) $V$-band light curve of {\target} from 2018 February
  to June (before MJD = 58300 days) and 2019 February to July 
  (after MJD = 58500 days).
Superoutburst (SO) is shown in detail in panel (b), and
  the three normal outbursts (NOs: N1, N2, and N3) are shown
  in panels (c), (d), and (e), respectively.
Color evolution (lower panels) for each of the four outbursts 
  is shown in panels (f), (g), (h), and (i) 
  for the SO, N1, N2, and N3, respectively.
Data with errors $< 0.2$ mag are used.
Short green vertical lines at the bottom of panel (a) show the epochs
  when the $V$-band observations were made.
Four blue star symbols in panel (a) denote the quiescent phase magnitudes
  obtained from photometry on stacked images of 10 exposures ($= 600$ seconds).
The mean of these quiescent phase magnitudes are denoted as blue horizontal
  dotted lines at $V = 22.59 \pm 0.18$ mag in panels (a) to (e).
While the gray dashed line in panel (b) shows the polynomial fitting result for the SO
  with order 8,
  we tried another polynomial fitting for the N2 (order 4, panel (d))
  and this fitting function is overplotted on N1 (panel (c)) and N3 (panel (e))
  for comparison.
The open circles in panels (f) to (i) are data points, 
  the dashed lines in panel (f) are polynomial fittings
  and those in panel (g) to (i) are mean colors.
The three star symbols on the left of panel (f) show the quiescent phase
  colors (see text for details).
}
\label{fig:fulllc}
\end{figure*}

Figure~\ref{fig:fulllc} (f) to (i) (lower panels) shows the color evolutions of
  \target, where open circles are the observed data points,
  dashed lines in panel (f) are polynomial fittings,
  and those in panels (g) to (i) are mean colors.
Figure~\ref{fig:fulllc} (f) shows that all the three colors remain constant
  ($B-V = 0.19 \pm 0.03$, $V-I = -0.31 \pm 0.03$, and
  $B-I = -0.12 \pm 0.03$ mag) from MJD $= 58205$ to 58215 days
  during which the outburst brightness changes.
As it gets closer to the end of the SO and to the quiescent phase magnitudes,
  all the colors get redder.
This reddening near the end of the decline can be explained by 
  the cooling of the accretion disk \citep{Cannizzo86,Cannizzo87}.
As a cooling front propagates inward, the $I$-band flux responds more sensitively 
  than the shorter-wavelength fluxes, causing 
  the colors containing the $I$-band get redder earlier than the $B-V$ color.
This effect arises because the contribution from 
  steady mass accretion, which is more prominent at longer wavelengths, 
  becomes increasingly significant as soon as 
  the outburst component, which is more evident at shorter wavelengths, begins to fade
  \citep{Hameury20}.

We calculate the quiescent phase colors 
  from the quiescent phase magnitudes obtained in \S\ref{sec:LC} 
  (average from Table~\ref{tab:quiescence_table})
  and show them as star symbols 
  in Figure~\ref{fig:fulllc} (f).
Just as in the case of KSP-OT-202104a \citep{kim26},
  which is a WZ Sge-type DN belonging to the SU UMa-type DNe,
  the SO colors converge to the quiescent phase colors
  as the outburst fades away.

The mean colors of the three NOs generally remain
  constant and are shown in Table~\ref{tab:meancolor_table}.
The mean $B-V$ colors are $0.16, 0.17,$ and $0.24$ mag,
  for N1, N2, and N3, respectively.
As in the SO, the colors of N1 and N2 (panels (g) and (h))
  show reddened values after the outbursts vanished
  and reached the quiescent phase.
All the three color indices show systematically bluer in N1
  and redder in N3 than in N2,
  although the offsets are small.

For the SU UMa-type DN KSP-OT-201611a, \citet{lee19} have shown
  that the color values are almost constant at $B-V = 0.23$ and $V-I = 0.06$ mag
  (extinction uncorrected) during the SO,
  while they show linear change during the NO
  ($B-V$ gets bluer and $V-I$ becomes redder). 
The nearly constant color values during the SOs of KSP-OT-201611a and our target
  are naturally explained by the constant viscosity model
  in the accretion disk, which predicts an approximately constant
  mass transfer rate \citep{lee19, Cannizzo87}.
Although the colors of {\target} N2 appear to remain nearly constant in 
  Figure~\ref{fig:fulllc} (h), slight color variations are still evident.
The $B−V$ color shows a slight tendency to become bluer after the peak, while 
  the $V−I$ color reaches its bluest value before the peak and 
  subsequently becomes progressively redder.
These trends are consistent with the color evolution described by \citet{lee19}
  for their NO.
During the NO, thermal instability triggers a transition of the outer accretion disk 
  from a cool state to a hot state.
The heating front then propagates inward, eventually bringing the entire accretion disk 
  into the hot state.
At this stage, the system reaches its bluest color \citep{Cannizzo86}.
Subsequently, the development of UV emission in the inner disk slightly delays 
  the occurrence of the bluest color in $B-V$.
In contrast, the bluest color is reached earlier in $V-I$, followed by 
  a more rapid evolution toward redder colors 
  due to the supply of cool material to the outer regions of the accretion disk.

% ------------------------------------------- Table1
\begin{deluxetable*}{c |cccc|cc}\label{tab:quiescence_table}
\tablecaption{Quiescent Phase Photometry of \target}
\tablehead{
& MJD & MJD & MJD & Number of & & \\
Band & Middle & First image & Last image & images & mag & error
}
\startdata
$B$ & 58172.2030 & 58170.2024 & 58174.2037 & 10 & 23.05 & 0.12 \\
$V$ & 58172.0643 & 58170.0604 & 58174.0682 & 10 & 22.60 & 0.12 \\
$I$ & 58172.2749 & 58170.4802 & 58174.0696 & 10 & 22.21 & 0.15 \\
\hline
$B$ & 58251.5324 & 58250.0719 & 58252.9930 & 10 & 22.71 & 0.08 \\
$V$ & 58251.7486 & 58250.7450 & 58252.7522 & 10 & 22.30 & 0.10 \\
$I$ & 58251.2673 & 58250.0748 & 58252.4599 & 10 & 21.95 & 0.11 \\
\hline
$B$ & 58573.0818 & 58571.1044 & 58575.0592 & 10 & 22.99 & 0.13 \\
$V$ & 58573.1298 & 58571.1990 & 58575.0606 & 10 & 22.74 & 0.15 \\
$I$ & 58573.4696 & 58571.9362 & 58575.0031 & 10 & 22.30 & 0.14 \\
\hline
$B$ & 58626.9017 & 58626.0328 & 58627.7707 & 10 & 22.77 & 0.09 \\
$V$ & 58626.8802 & 58626.0341 & 58627.7262 & 10 & 22.73 & 0.11 \\
$I$ & 58626.8816 & 58626.0355 & 58627.7277 & 10 & 22.25 & 0.10 \\
\enddata
\end{deluxetable*}

% ------------------------------------------- Table2
\begin{deluxetable*}{ccccc}\label{tab:meancolor_table}
\tablecaption{Mean Colors of the Outbursts of \target}
\tablehead{
Color & Superoutburst & N1 & N2 & N3 \\
& [mag] & [mag] & [mag] & [mag] 
}
\startdata
$B-V$ & $+0.19 \pm 0.03$ & $+0.16\pm0.08$ & $+0.17\pm0.05$ & $+0.24\pm0.08$ \\
$V-I$ & $-0.31 \pm 0.03$ & $-0.34\pm0.04$ & $-0.30\pm0.06$ & $-0.27\pm0.07$ \\
$B-I$ & $-0.12 \pm 0.03$ & $-0.18\pm0.04$ & $-0.12\pm0.06$ & $-0.06\pm0.07$ \\
\enddata
\end{deluxetable*}

%------------------------------------------------------------------------------
\subsection{Period Analysis} \label{sec:period}
During SOs of SU UMa-type DNe, small amplitude (up to 0.2 mag) magnitude oscillations 
  known as `superhumps' occur. 
These oscillations are caused by varying tidal dissipation 
  in a precessing, eccentric and tidally distorted accretion disk
  \citep[e.g.,][]{Kato09_S395}. 
SU UMa-type DNe exhibit three stages during the superhump phase:
 `A' is an early evolutionary phase with a longer superhump period,
 `B' is an intermediate phase with systematically varying periods, and
 `C' is a phase with shorter and almost constant period \citep{Kato09_S395}.
To obtain the period of the superhump ($P_{\rm sh}$) during the plateau
  of the SO, we use the phase dispersion minimization (PDM: \citealt{Stellingwerf78})
  method, which finds periods from unevenly sampled time-series data
  \citep{Tampo26, kim26}.
Due to the relatively low cadence of our observations, it is difficult 
  to directly identify the superhump stages using an O−C diagram.
However, empirical studies have shown that the transition from superhump 
  stage A to stage B generally occurs a few days after the outburst peak 
  \citep{Patterson11,Kato09_S395}.
Since we measured the superhump period of {\target} using observations 
  obtained at MJD $58210-58217$, approximately 3.5 days after the $V$-band peak, 
  we consider the measured superhumps to be approximately in stage B.

\begin{figure*}   % Fig 3 ----------------------
\epsscale{0.9}
\plotone{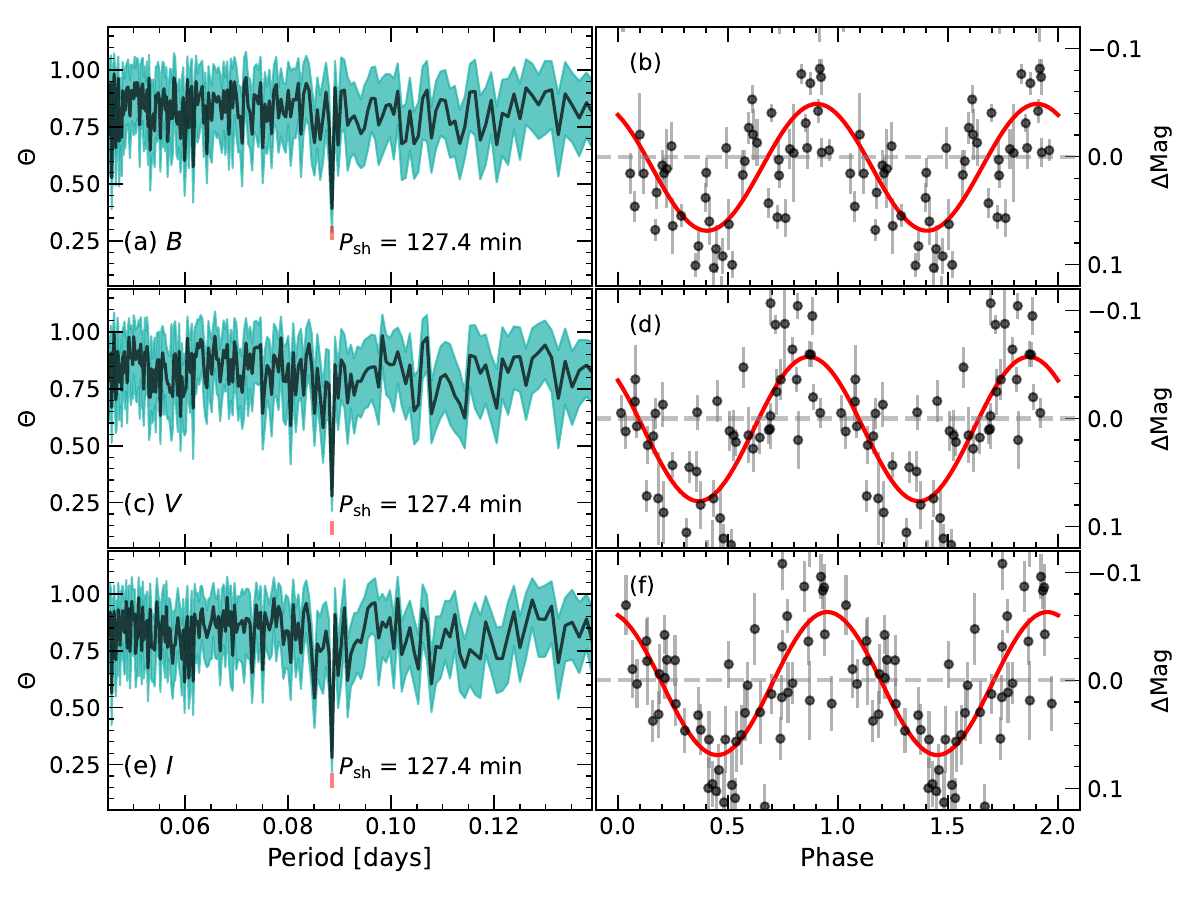}
\caption{Results of the PDM analysis (left) and phase-folded profiles (right)
  of \target.
Panels (a) and (b) are for $B$-band,
  (c) and (d) are for $V$-band, and
  (e) and (f) are for $I$-band.
In the left panels, the black solid lines are median $\Theta$ and
  the shaded areas show 90\% confidence ranges from 1,000 bootstrap resamplings.
All the $B, V,$ and $I$-band data show the same superhump period of 
  127.4 minutes 
  ($=2.123$ hr $=0.08847$ days).
Red solid lines on the right panels are sine curves fitted to 
  the data with the obtained period.
}
\label{fig:PDM}
\end{figure*}

Figure~\ref{fig:PDM} shows the results of the PDM analysis.
The black lines on the left panels are the median $\Theta$ and
  the shaded areas are the 90\% confidence ranges
  from 1,000 bootstrap resamplings.
All of the three ($B, V,$ and $I$) bands show the same obvious peaks
  at $P_{\rm sh} = 127.4$ minutes ($=2.123$ hr $=0.08847$ days).
The right panels are phase-folded profiles drawn using this period
  with the sine curves fitted to the data shown in red solid lines.
Considering the fact that the superhump periods of SU UMa-type DNe are 
  usually a few percent longer than the orbital periods
  \citep[``positive superhumps'';][]{patterson05,kato17ninth,kim26},
  the orbital period of \target\ would be 
  a few percent shorter than the superhump period derived here.

%==============================================================================
\section{Discussion} \label{sec:discussion}

\subsection{Period Gap}  \label{sec:Pgap}
The period gap is characterized by a deficit of CVs with orbital periods 
  between $2-3$ hours \citep{Rappaport83,Spruit83},
  while there is a sharp cutoff at the lowest value
  known as the `period minimum' of $\sim 76$ minutes
  ($70-80$ min: \citealt{knigge06,kim26}). 
To explain the period gap, the `disrupted magnetic braking' model was 
  proposed \citep{Rappaport83,Spruit83}.
The values of the upper and lower edges of the period gap
  are not known exactly: 
  \citet{knigge06} and \citet{knigge11} claim
    $P_{\rm gap,-} = 2.15 \pm 0.03$ h and  % 129 m
    $P_{\rm gap,+} = 3.18 \pm 0.04$ h, and % 191 m
  \citet{Schreiber24} suggests
    $P_{\rm gap,-} = 2.45$ h and % 147 m
    $P_{\rm gap,+} = 3.18$ h. % 191 m
It might also be possible that there would not be
  sharp boundaries for the period gap \citep{Schreiber24}.
Considering the fact that usually the superhump period is a few percent
  longer than the orbital period (``positive superhumps''; \citealt{Osaki03, kato09_61_601,kim26}),
  it is probable that the orbital period of {\target} might be 
  a bit shorter than $P_{\rm gap,-}$.
This means that {\target} could be, if the magnetic braking model is correct,
  an object that just resumed the mass transfer and became observable to us.
Better understanding in the future would be made
  on the changes taking place at this transition stage
  if enough observations are made for the objects near the edge.

%------------------------------------------------------------------------------
\subsection{Gaia Observations}  \label{sec:GAIA}
We have searched the archival data of
  the American Association of Variable Star Observers, % (AAVSO)
  Catalina Real-Time Transient Survey \citep{Drake09_CRTS}, %(CRTS)
  All-Sky Automated Survey for SuperNovae \citep{Kochanek17_ASASSN}, % ASASSN
  and Zwicky Transient Facility (ZTF, \citealt{Bellm19_ZTF}), % Masci19_ZTF
  % https://www.ztf.caltech.edu/
  to see if any of them contain signals from \target,
  but could not find any.

We searched the Gaia DR3 catalog \citep{Gaia23} for distance information to \target\
  but could not obtain a distance estimate. 
However, we found that \target\ had been observed six times by Gaia
  as shown in Table~\ref{tab:Gaia_table}.
The first three observations (mean $G \approx 17.86$ mag) correspond 
  to $\sim 3$ days after the peak of SO in Figure~\ref{fig:fulllc} (b).
The detection in 2021 at $G = 17.58$ mag appears to be made at 
  another SO, while those in 2020 could be during a NO
  or another SO in rising or declining phases.
Although the Gaia observations spanned a total of 3 years and 2 months, 
  \target\ was detected only during outbursts because 
  it was too faint during quiescence.
Future observations with larger telescopes (e.g., Rubin/LSST) 
  may enable us to investigate its recurrence cycle.

% ------------------------------------------- Table3
\begin{deluxetable}{ccc}\label{tab:Gaia_table}
\tablecaption{Gaia Observations for \target\ (ID: Gaia18aus)}
\tablehead{
MJD (days) & Date (UT) & G (mag)
}
\startdata
58209.16558 & 2018 Apr 1 $03^{h}$ $58^{m}$ $26.1^{s}$ & 17.86 \\
58209.41574 & 2018 Apr 1 $09^{h}$ $58^{m}$ $39.9^{s}$ & 17.82 \\
58209.48975 & 2018 Apr 1 $11^{h}$ $45^{m}$ $14.4^{s}$ & 17.91 \\
\hline
59026.63265 & 2020 Jun 26 $15^{h}$ $11^{m}$ $01.0^{s}$ & 18.24 \\
59026.70666 & 2020 Jun 26 $16^{h}$ $57^{m}$ $35.4^{s}$ & 18.31 \\
\hline
59372.19819 & 2021 Jun 7 $04^{h}$ $45^{m}$ $23.6^{s}$ & 17.58 \\
\enddata
\end{deluxetable}

%------------------------------------------------------------------------------
\subsection{Morphology of the Outburst LCs}  \label{sec:Others}
\begin{figure*}   % Fig 5 ------------------------
\epsscale{0.8}
\plotone{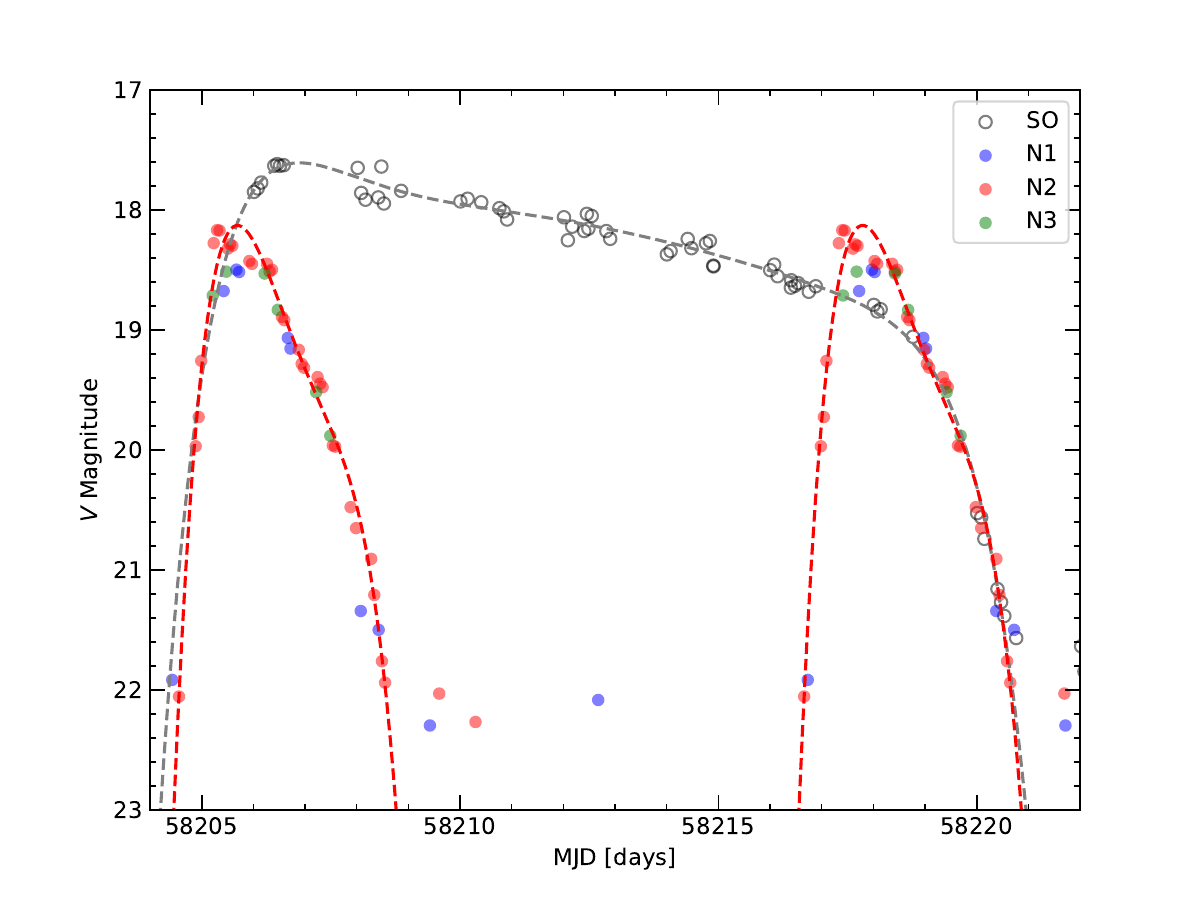}
\caption{Overlap of the outburst light curves. 
$V$-band SO is shown in gray open circles and gray dashed line.
Three normal outbursts in $V$-band (N1 blue, N2 red, and N3 green)
  are plotted as filled circles and red dashed lines
  twice at rising and declining parts.
}
\label{fig:LCoverlap}
\end{figure*}

We overplotted the outburst LCs in Figure~\ref{fig:LCoverlap} :
  SO $V$-band data are shown in gray color (open circles and dashed line), 
  three NOs in filled circles (N1 blue, N2 red, and N3 green) and 
  red dashed lines.
The three NOs are shown twice at the rising and declining parts
  of the SO.
The three NOs match very well with each other and with the SO
  at the rising and declining parts,
  while there are some scattered points near the NO peaks
  implying a bit different peak brightnesses.
Note that there are not enough data points at the rising part of the SO,
  therefore this part should not be regarded to be reliable.

For the 21 outbursts of the U Gem-type DN CH UMa during $1973-1998$,
  \citet{Simon00_354} showed that their peak brightnesses are bimodal:
  ``bright'' and ``faint'' being separated at 12 $\rm mag_{vis}$.
While the {\it bright} outbursts 
  are classified into two types of fast and slow decays % their Fig 2
  by the decay branches, 
  the {\it bright} outbursts are symmetric and % their Fig 2
  the {\it faint} outbursts are asymmetric     % their Fig 3
  (decay being slower than the rise).
On the other hand, \citet{Bath83} classified the outburst LCs
  of SS Cygni from 1897 till 1979
  into four types : A, B, C, and D from the steepest rise 
  to the most shallow rise patterns.
Outbursts propagating outward from the inner disk (‘inside-out’) 
  produce symmetric LCs, whereas 
  inward-propagating outbursts from the outer disk (‘outside-in’) 
  yield asymmetric LCs with faster rises than decays
  \citep{Cannizzo86,Cannizzo87,lee19}. 
The asymmetric LC in Figure~\ref{fig:LCoverlap},
  characterized by steeper rises than decays,
  supports the outside-in scenario as the more likely mechanism
  for the $2018-2019$ outbursts of \target.
The NOs in SU UMa systems are generally believed to be outside-in type 
  \citep{Buat01},
  and outside-in outbursts appear to be more common than inside-out outbursts 
  in numerous studies of individual SU UMa-type systems 
  \citep[see, e.g.,][]{Bruch26}.

%==============================================================================
\section{Summary and Conclusion} \label{sec:summary}

In this work, from two years of observations by the KMTNet Supernova Program
  for the SU UMa-type DN \target, we have studied the LCs and obtained the results
  as summarized in the following.
\begin{itemize}
\item 
One superoutburst (SO) and three normal outbursts (NOs) are identified 
  during the observations of $2018 − 2019$.

\item By stacking ten images of the quiescent phase, we estimate 
  the quiescent phase magnitude of $V = 22.59 \pm 0.18$ mag.

\item The LC parameters of SO are : magnitude amplitude of $4.97 \pm 0.18$ mag,
  peak magnitude of $V = 17.62 \pm 0.02$ mag, and 
  outburst duration of 16.59 days.
  The second NO (N2), among the three NOs, is most thoroughly observed
  and gives the following LC parameters : amplitude $4.46 \pm 0.18$ mag, 
  peak magnitude $V \approx 18.13$ mag, and duration of 4.18 days.

\item Applying the phase dispersion minimization method 
  to the superhump data during the plateau phase of the SO yields a 
  superhump period of $P_{\rm sh} = 127.4$ minutes 
  ($=2.123$ hr $=0.08847$ days).
We assume the orbital period of \target\ would be a few percent shorter than
  this superhump period.
\citet{Schreiber24} gives the lower and upper boundaries 
  of the orbital period gap to be $P_{\rm gap,-} = 2.45$ hours and
  $P_{\rm gap,+} =3.18$ hours, respectively.
It might be possible that \target\ just crossed the lower boundary
  meaning the system recently resumed the mass transfer after the detached phase
  invisible to us for a while.

\item The asymmetric shapes of the outburst LCs of \target\
  support the outside-in propagation scenario 
  where the outburst instability started at the outer part of the accretion disk
  and moved inward.
\end{itemize}

%\clearpage

%==============================================================================
\begin{acknowledgments}
We thank the referee for the very helpful comments that improved the manuscript.
This research has made use of the KMTNet system operated by 
  the Korea Astronomy and Space Science Institute (KASI) at three host sites 
  of CTIO in Chile, SAAO in South Africa, and SSO in Australia.
Data transfer from the host site to KASI was supported by 
  the Korea Research Environment Open NETwork (KREONET).
This research was supported by the Korea Astronomy and Space Science Institute 
  under the R\&D program (Project No. 2026-1-860-00) supervised by 
  the Ministry of Science and ICT.
This work has made use of data from the European Space Agency (ESA) mission 
  Gaia (https://www.cosmos.esa.int/gaia), processed by the Gaia Data Processing 
  and Analysis Consortium (DPAC, https://www.cosmos.esa.int/web/gaia/dpac/consortium).
Funding for the DPAC has been provided by national institutions, in particular 
  the institutions participating in the Gaia Multilateral Agreement.
\end{acknowledgments}

%\software{KCORRECT v4.2 \citep{Bla07}}

\bibliography{draft}

\end{document}